\documentclass[twocolumn,floatfix,prb,aps,showpacs]{revtex4-1}
\usepackage{graphicx,amsmath,amssymb,color}
\usepackage{nicefrac,xcolor}
\usepackage{multirow, makecell, ragged2e}
\usepackage[titletoc,title]{appendix}
\usepackage{ulem}
\usepackage{diagbox}

\newcommand{\be}{\begin{equation}}
\newcommand{\ee}{\end{equation}}

\newcommand{\ba}{\begin{eqnarray}}
\newcommand{\ea}{\end{eqnarray}}

\renewcommand{\vec}[1]{\mbox{\boldmath$#1$}}

\def\beq{\begin{equation}}
\def\eeq{\end{equation}}

\begin{document}

\title{Interplay between fractional quantum Hall liquid and crystal phases at low filling}
\author{Zheng-Wei Zuo$^{1,2,\dagger}$, Ajit C. Balram$^{3,\dagger}$, Songyang Pu$^{1}$, Jianyun Zhao$^{1}$, Thierry Jolicoeur$^{4}$, A. W\'ojs$^{5}$, and J. K. Jain$^{1}$}
\affiliation{$^1$Department of Physics, The Pennsylvania State University, University Park, Pennsylvania 16802, USA}
\affiliation{$^2$School of Physics and Engineering, and Henan Key Laboratory of Photoelectric Energy Storage Materials and Applications, Henan University of Science and Technology, Luoyang 471023, China}
\affiliation{$^{3}$Institute of Mathematical Sciences, HBNI, CIT Campus, Chennai 600113, India}
\affiliation{$^{4}$Institut de Physique Th\'eorique, Universit\'e Paris-Saclay, CNRS, CEA, 91190 Gif sur Yvette, France}
\affiliation{$^{5}$Department of Theoretical Physics, Wroc\l{}aw University of Science and Technology, 50-370 Wroc\l{}aw, Poland}  
\date{\today}

\begin{abstract}
The nature of the state at low Landau-level filling factors has been a longstanding puzzle in the field of the fractional quantum Hall effect. While theoretical calculations suggest that a crystal is favored at filling factors $\nu\lesssim 1/6$, experiments show, at somewhat elevated temperatures, minima in the longitudinal resistance that are associated with fractional quantum Hall effect at $\nu=$ 1/7, 2/11, 2/13, 3/17, 3/19, 1/9, 2/15 and 2/17, which belong to the standard sequences $\nu=n/(6n\pm 1)$ and $\nu=n/(8n\pm 1)$. To address this paradox, we investigate the nature of some of the low-$\nu$ states, specifically $\nu=1/7$, $2/13$, and $1/9$, by variational Monte Carlo, density matrix renormalization group, and exact diagonalization methods. We conclude that in the thermodynamic limit, these are likely to be incompressible fractional quantum Hall liquids, albeit with strong short-range crystalline correlations. This suggests a natural explanation for the experimentally observed behavior and a rich phase diagram that admits, in the low-disorder limit,  a multitude of crystal-FQHE liquid transitions as the filling factor is reduced.
\end{abstract}

\maketitle

\section{Introduction}
The physics of the fractional quantum Hall effect (FQHE), right from its beginning~\cite{Tsui82, Laughlin83}, has been intertwined with the physics of the expected Wigner crystal phase in the lowest Landau level (LLL)~\cite{Wigner34, Lozovik75}. There has been a great deal of theoretical~\cite{Maki83,Lam84,Levesque84,Yi98,Yang01,Narevich01,Chang05,Archer13,Zhao18} and experimental~\cite{Shayegan07,Fertig07,Jiang90, Goldman90,Williams91, Paalanen92, Santos92,Santos92b,Manoharan94a, Engel97, Pan02, Li00, Ye02, Chen04, Csathy05, Pan05,Sambandamurthy06, Chen06,Liu14a,Zhang15c,Deng16,Jang17, Chen18a} work addressing this issue, and the following picture is widely accepted: (i) At filling factors $\nu=n/(2n\pm 1)$, $\nu=n/(4n\pm 1)$, where $n$ is a positive integer, and their hole partners, the ground state is a FQH liquid. These states are understood as integer quantum Hall (IQH) state of composite fermions (CFs)~\cite{Jain89}, which are bound states of electrons and an even number of quantized vortices. (ii) In $n$-doped GaAs quantum wells, an insulating phase is seen between $\nu=1/5$ and $\nu=2/9$ and also below $\nu=1/5$~\cite{Jiang90, Goldman90}. A strong case may be made that these insulators represent a crystal pinned by the disorder. Similarly, insulating states, seen in the vicinity of $\nu=1/3$~\cite{Santos92, Santos92b} are also viewed as disorder-pinned crystals. (iii) Theoretically, it has been demonstrated that the crystal of composite fermions~\cite{Chang05, Archer13, Zhao18} is energetically better than the crystal of electrons. The number of vortices bound to composite fermions is fewer than the maximum number of available vortices, which leaves composite fermions with enough freedom to form a crystal. (For example, at 1/7 and 1/9, the best crystals have 4 and 6 vortices bound to composite fermions.) In particular, the crystal formed in between 1/5 and 2/9 is explained as a crystal of $^2$CFs~\cite{Archer13}. (The symbol $^{2p}$CF denotes composite fermions carrying $2p$ vortices.)
(iv) Finally, the FQHE terminates for $\nu\lesssim 1/6$, where the crystal phase dominates. 

It is the last assertion (iv) that we address in this paper. The motivation is as follows. Experiments clearly show that the state for $\nu<1/5$ is insulating with exponentially high resistance at the lowest temperatures. At the same time, signatures of FQHE at $\nu=1/7$ and $2/11$ have been reported by Goldman {\it et al.}~\cite{Goldman88} and Mallett {\it et al.}~\cite{Mallett88}, respectively. Moreover, Pan {\it et al.}~\cite{Pan02,Pan08} have observed developing FQH states at $\nu=$ 1/7, 2/11, 2/13, 3/17, 3/19, 1/9, 2/15, and 2/17 at elevated temperatures (see Table I of Ref.~\cite{Pan08} for a list of observed fractions), which belong to the standard Jain sequences at $\nu=n/(6n\pm 1)$ and $\nu=n/(8n\pm 1)$ arising from the integer quantum Hall effect of $^6$CFs and $^8$CFs. These observations are not readily reconcilable with the assertion (iv).

We therefore revisit the generally accepted view that the region below $\nu\lesssim 1/6$ is dominated by the crystal phase. The issue is ultimately an energetic one and requires an accurate understanding of both the liquid and the crystal phases. We find that the competition between the FQHE liquid and the crystal phases is much subtler than previously believed. On the whole, our calculations support, in an ideal disorder-free situation, an incompressible FQHE liquid with strong short-range crystalline correlations at fractions such as $\nu=1/7$, 2/13, 1/9, which belong to the $\nu=n/(6n\pm 1)$ and $\nu=n/(8n\pm 1)$ sequences.

We provide here a summary of our results obtained from three different methods we apply to this problem. Throughout this work, we shall assume that the external magnetic field is strong enough to fully spin polarize the electrons. Furthermore, we shall consider an ideal system with zero width, zero Landau-level (LL) mixing, and no disorder.

\underline{Variational Monte Carlo (VMC):} 
The statement that the crystal is stabilized for $\nu\lesssim 1/6$ is largely based on variational comparisons between candidate FQHE and crystal states. The validity of the variational studies, however, depends on the accuracy of the wave functions used in the study. For the crystal phase, the CF crystal (CFC) wave functions are extremely accurate at low filling factors~\cite{Chang05}. In particular, the $^4$CF crystal has the lowest energy at $\nu=1/7$, and the $^6$CF crystal has the lowest energy at $\nu=1/9$.  We find that the best CF crystal has lower energy than the ``bare" Laughlin/Jain wave functions~\cite{Laughlin83, Jain89} for $\nu=1/7$, $2/13$ and $1/9$. However, the bare Laughlin/Jain wave functions are not as accurate representations of the liquid at small fillings~\cite{Mandal03, Chang05} as they are for $\nu=n/(2n\pm 1)$. The variational energy of the liquid state at low filling factors can be significantly improved by allowing a renormalization of composite fermions through dressing by CF excitons, using the method of CF diagonalization (CFD)~\cite{Mandal02}. We find that after such renormalization, the energies of the CF liquids become competitive with those of the best CF crystal. 

In this context, we note that an earlier work~\cite{Kamilla97c} found an excitonic instability at $\nu=1/9$ and lower fillings within the bare CF theory. However, this instability was later found to disappear with improved wave functions for the ground and excited states obtained using CFD~\cite{Peterson03}.

\underline{Density matrix renormalization group (DMRG):} Earlier calculations in the torus~\cite{Yang01} and disk~\cite{Chang05} geometries favored a crystal phase at $\nu=1/7$ and $\nu=1/9$. However, these calculations were performed for $N=6$ and thus are not necessarily indicative of the behavior in the thermodynamic limit. As an example, it was shown in Ref.~\cite{Chang06} that the crystal state wins over the liquid even for $\nu=1/5$ for small systems, and it is necessary to go to sufficiently large systems (with $N\geq 10$) to see that the actual state is an FQHE liquid. To see if analogous physics might be present at $\nu=1/7$ and $\nu=1/9$, we have performed extensive DMRG studies in the cylindrical geometry by considering systems with up to $N=25$ particles. On a cylinder, some particle numbers are more favorable to crystal formation, because the geometry accommodates a triangular crystal for these particle numbers. We find that for $\nu=1/3$ the system is clearly a liquid, independent of $N$. For other fractions, we find liquid-like correlations for particle numbers ($N=6$, 12, and 20) for which a triangular crystal cannot be accommodated on the cylinder. For particle numbers that favor a crystal ($N=9$, 16, and 25), we find strong crystalline correlations. While our calculations do not decisively discriminate between the liquid and crystal phases in the thermodynamic limit, they do not rule out, for $\nu=1/5$, 1/7 and 1/9, an FQHE liquid with strong short-range crystalline correlations.

\underline{Exact diagonalization (ED):} We perform extensive ED studies in the spherical geometry, going to much larger systems than before. In our studies, we find that the states at $\nu=1/7$, 2/13, 1/9, and 1/11 appear to be incompressible FQH liquid states by all measures. They have uniform ground states, i.e. have total orbital angular momentum $L=0$ in the spherical geometry, which is an important property of incompressible states (for other states, in general, $L>0$). They have the expected excitation spectrum and have robust charge and neutral gaps that extrapolate to non-zero values in the thermodynamic limit. The ground states have a significant overlap with the Laughlin or Jain wave functions. For $N=10$ particles, the exact energy as a function of flux also shows downward cusps at the special filling factors $\nu=1/7$ and $2/13$. The study in the spherical geometry fully supports an FQHE liquid at the low filling factors we have studied. Even though the spherical geometry disfavors a crystal, because a triangular crystal cannot be perfectly accommodated on the surface of a sphere, it does provide ample freedom to signal a departure from FQHE by producing either a non-uniform ground state with $L\neq 0$ or a liquid that is not described by the standard FQHE physics. We therefore take the ED results as providing nontrivial support for the stabilization of FQHE states at low filling factors.
 
Based on these considerations, especially the ED results in the spherical geometry, we conclude that, overall, our calculations favor an FQHE liquid at $\nu=1/7$ and 1/9, and presumably at certain other filling factors of the form $\nu=n/(6n\pm 1)$ and $\nu=n/(8n\pm 1)$, in the thermodynamic limit. This has clear experimental consequences, which ought to be testable in better quality samples; it is worth mentioning that a significant jump in the mobility has recently been achieved in GaAs based two dimensional systems~\cite{Chung20,Pfeiffer20}.

We note that signatures of FQHE at $\nu=1/7$ in a higher LL have recently been reported in WSe$_{2}$~\cite{Shi19}. Our present study is confined to the LLL filling factors.

The plan for the rest of the paper is as follows. We first provide a primer on composite fermion theory and spherical geometry in Sec.~\ref{sec: primer}. Results obtained from variational Monte Carlo calculations on the sphere using the CF wave functions are given in Sec.~\ref{sec: VMC}. Section~\ref{sec: DMRG} contains results from DMRG calculations. In Section~\ref{subsec: EDresults} we provide results obtained from our extensive exact diagonalization studies on the sphere. Section~\ref{sec: discussion} discusses the relation of our results to experiments and suggests a schematic phase diagram for FQHE at low filling factors. 

\section{Primer on spherical geometry and composite fermion theory}
\label{sec: primer}

Our exact diagonalization and variational Monte Carlo calculations are carried out on the Haldane sphere~\cite{Haldane83}, where $N$ electrons are confined to the surface of a sphere in the presence of a radial magnetic flux of $2Qhc/e$ ($2Q$ is an integer) generated by a magnetic monopole placed at the center of the sphere. The radius of the sphere is given by $R=\sqrt{Q}\ell$, where $\ell=\sqrt{\hbar c/eB}$ is the magnetic length and $B$ is the perpendicular magnetic field. The quantity $\phi_0=hc/e$ is called the flux quantum.  Appropriate to this geometry, the total orbital angular momentum $L$ and its $z$-component $L_{z}$ are good quantum numbers. The single-particle eigenstates for a given $2Q$ are labeled by the single-particle angular momentum $l=|Q|$ and its z-component $l_z=-|Q|, \cdots |Q|$; these are referred to as orbitals. Incompressible quantum Hall states at filling factor $\nu$ occur on the sphere when $2Q=N/\nu-\mathcal{S}$, where $\mathcal{S}$ is a topological quantum number called the shift~\cite{Wen92}. These states are distinguished by the fact that (i) they are uniform, i.e. have $L=0$, for all $N$, and (ii) they have a non-zero gap to excitations in the thermodynamic limit. These are the criteria that we will use to ascertain whether the actual state is incompressible. In contrast, a compressible state, in general, has $L\neq 0$ and no well-defined gap.

The  phenomenology of FQHE occurring in the LLL is understood using the CF theory~\cite{Jain89,Jain07}, which maps strongly interacting electrons at filling factor $\nu = \nu^{*}/(2p\nu^{*}+ 1)$ into weakly interacting CFs carrying $2p$ vortices at filling factor $\nu^{*}$. One of the consequences of this mapping is that when $\nu^{*}=n$, where $n$ is a positive integer, an incompressible FQH state of electrons occurs at filling factor $\nu=n/(2pn+1)$. The Jain CF wave function for the incompressible FQH ground state at $\nu=n/(2pn+1)$ is given by
\begin{equation}
\Psi^{\rm CF}_{n/(2pn+1)} = {\cal P}_{\rm LLL}\Phi_n \Phi_1^{2p},
\label{eq_CF_wfs}
\end{equation}
where $\Phi_n$ is the IQH state constructed at the effective magnetic monopole strength $2Q^* = N/n-n$. The symbol ${\cal P}_{\rm LLL}$ represents projection into the LLL, for which we use the Jain-Kamilla (JK) method~\cite{Jain97,Jain97b,Jain07,Moller05,Davenport12,Balram15a}. Because the shifts add, the incompressible state at $\nu=n/(2pn+1)$ is predicted to occur at shift $\mathcal{S}=n+2p$. The above wave function reduces to the Laughlin wave function for $\nu=1/(2p+1)$. The wave function of the ground state as well as the excitations obtained from the CF theory accurately capture the corresponding true Coulomb states in the LLL~\cite{Dev92,Jain97,Peterson03,Moller05,Jain07,Majumder09,Balram13,Jain14,Majumder14,Balram16b}. 

Similarly, using the analogy to the IQH effect, wave functions for the low-energy excitations of the FQH state, namely quasiparticles and quasiholes (which are created upon removal or insertion of flux quanta in the ground state) can also be constructed. In particular, CF theory predicts that the total orbital angular momentum of the single quasihole (QH) or single quasiparticle (QP) state at $\nu=n/(2pn+1)$ is $L^{\rm QH}=(N+(n-1)^{2})/(2n)$ and $L^{\rm QP}=(N+n^{2}-1)/(2n)$. In particular, the single quasihole or single quasiparticle state at $\nu=1/(2p+1)$ is obtained respectively by the addition or removal of a single flux quantum $hc/e$ in the ground state and occurs at $L^{\rm QH}=N/2=L^{\rm QP}$.

The Laughlin state at $\nu=1/(2p+1)$~\cite{Laughlin83} is the unique densest exact zero-energy ground state of the short-range interaction specified by the two-body Haldane pseudopotentials $V_{m}=1~\forall m\leq 2p$ and $V_{m}=0~\forall m\geq 2p+1$~\cite{Haldane83}. Here $V_{m}$ is the energy of a pair of electrons in the relative angular momentum state $m$. No local interactions in the LLL are known which produce the wave functions of Eq.~(\ref{eq_CF_wfs}) at other fractions as exact zero-energy ground states~\cite{Sreejith18}. 

Next, we describe the various electron and composite fermion crystal states considered in this work. Crystal states can occur when the electrons or composite fermions prefer to occupy localized wave packets to minimize the strong repulsion of the Coulomb interaction. In the spherical geometry, a wave packet localized at spinor coordinates $(U,V)$ is created by $\left(u U^{*} + v V^{*} \right)^{2Q}$, where $(u,v)$ are particle coordinates. The $2p$-vortex composite fermion crystal ($^{2p}$CFC) wave function is given by
\begin{equation}
 \Psi^{^{2p}{\rm CFC}}_{2Q} = \prod_{i<j}\left(u_{i}v_{j}-u_{j}v_{i}\right)^{2p}
\Psi^{\rm EC}_{2Q^*},
 \label{eq_crystal_wfs}
\end{equation}
\be
\Psi^{\rm EC}_{2Q^*}= {\rm Det}\left[ \left(u_{i} U_{j}^{*} + v_{i} V_{j}^{*} \right)^{2Q^{*}} \right],
\ee
where ${\rm Det}$ stands for determinant. Here $\Psi^{\rm EC}_{2Q^*}$ is the electron crystal, and $\Psi^{^{2p}{\rm CFC}}_{2Q}$ is a crystal of composite fermions because the factor $\prod_{i<j}\left(u_{i}v_{j}-u_{j}v_{i}\right)^{2p}$ attaches $2p$ vortices to each electron in the electron crystal. We have $2Q^{*}=2Q-2p(N-1)$ and only those values of $p$ are allowed that lead to a positive value of $2Q^{*}$. We treat $2p$ as a variational parameter. The crystal sites can also be thought of as variational parameters. 
We choose the crystal sites to lie at the Thomson locations~\cite{Thomson04,Thomson,Wales06,Wales09}, which minimizes the classical Coulomb repulsion energy of point particles on a sphere. Note that a triangular crystal, which is what the electrons organize themselves into in two dimensions to minimize the Coulomb repulsion energy, in the spherical geometry necessarily has some defects.

We note that for $\nu=1$, i.e., for $2Q^*=N-1$, we have $\Psi^{\rm EC}_{2Q^*}=\prod_{i<j}\left(u_{i}v_{j}-u_{j}v_{i}\right)$ (apart from normalization) independent of $(U_j,V_j)$, because that is the only wave function available within the LLL space. For that reason, $\Psi^{^{2p}{\rm CFC}}_{2Q}$ is identical to the Laughlin wave function for $\nu=1/(2p+1)$.

One quantity of interest is the pair-correlation function, which is defined as
\begin{equation}
g\left(  \mathbf{r}, \mathbf{r}^{\prime}\right)  ={1\over \rho_0^2}\left\langle \sum_{i\neq j}\delta\left(  \mathbf{r}-\mathbf{r}_{i}\right)\delta\left(  \mathbf{r}^{\prime}-\mathbf{r}_{j}\right)  \right\rangle, \label{eq:gr}
\end{equation} 
where $\rho_0$ is the average density. 
For a translationally invariant system, and in particular for the spherical geometry, we have $g\left(  \mathbf{r}, \mathbf{r}^{\prime}\right)=g\left( | \mathbf{r}-\mathbf{r}^{\prime}\right|)$. The pair-correlation function is essentially the density at $\vec{r}$ when one particle is fixed at $\vec{r}'$, normalized so that $g(| \mathbf{r}-\mathbf{r}^{\prime}|)\rightarrow 1$ in the limit $| \mathbf{r}-\mathbf{r}^{\prime}|\rightarrow \infty$ for a liquid.

\section{Variational Monte Carlo studies}
\label{sec: VMC}

Using the wave functions given in Eq.~(\ref{eq_CF_wfs}) we calculate the Coulomb energy of the FQHE liquid states at $\nu=1/7,~2/13$ and $1/9$ for large systems using the Metropolis Monte Carlo method~\cite{Binder10}. The ``bare'' Laughlin/Jain wave functions of Eq.~(\ref{eq_CF_wfs}) do not give the best description of the liquid ground state at the low fillings considered in our work~\cite{Mandal03}. The liquid state energies are significantly lowered by dressing the bare state with CF excitons. The energy of this renormalized ground state is calculated by the method of CFD~\cite{Mandal02, Jain07}. We quote energies of the liquid state at $\nu=1/7,~2/13$, and $1/9$ obtained by renormalizing the Laughlin/Jain state by dressing it with up to three CF excitons. In Table~\ref{tab:energies_exact_liquids_crystals_small_N} we show results for small systems where we compare the liquid energies against exact diagonalization. These correlation energies include contributions of the electron-background and background-background interactions. All the energies are quoted in standard Coulomb units of $e^2/(\epsilon\ell)$, where $\epsilon$ is the dielectric constant of the background host material.
We also evaluate the Coulomb energies of various crystal states of Eq.~(\ref{eq_crystal_wfs}) using the Monte Carlo method. As seen in Table~\ref{tab:energies_exact_liquids_crystals_small_N}, for $\nu=1/7$ and 2/13, the $^4$CF crystal has the lowest energy, whereas at $\nu=1/9$ the $^6$CF crystal has the lowest energy.

The primary conclusion is that the dressed liquid energies and the CF crystal energies are very close to the exact energies for all systems considered, and thus VMC cannot discriminate between the two.

\begin{table*}
\centering
\scalebox{0.8}{
\begin{tabular}{|c|c|c|c|c|c|c|c|c|c|c|c|c|}
\hline
$\nu$ & $N$ & $2Q$ &  $dim_{L_{z}=0}$ &  $dim_{L=0}$ & exact & \multicolumn{2}{|c|}{liquid energies ($e^2/(\epsilon\ell)$)} & \multicolumn{4}{|c|}{crystal energies ($e^2/(\epsilon\ell)$)}  & overlap\\ \hline
      & & & & & &	  bare		& \multicolumn{1}{|c|}{renormalized} 			&	electron crystal  	& \multicolumn{3}{|c|}{CF crystal}		&	\\ \hline
      & & & & &	  &	Laughlin/Jain	& up to 3 CF excitons     			&			        & $^{2}$CF	& $^{4}$CF 	& $^{6}$CF 			&	$|\langle\psi^{\rm exact}|\psi^{\rm L/J}\rangle|$	\\ \hline
1/7 & 4	& 21	&241		&4		& -0.279799 &-0.279320(0)   &-0.27980(0)&-0.27366(5)&-0.27671(8)&-0.27814(5)& -0.279320(0)&0.9741\\ \hline
    & 5 & 28    &2,649		&7		& -0.279223 &-0.279157(0)   &-0.27920(1)&-0.27198(7)&-0.27546(8)&-0.27729(7)& -0.279157(0)&0.9972\\ \hline
    & 6 & 35    &32,134		&47		& -0.280706 &-0.279773(0)   &-0.28069(0)&-0.27576(0)&-0.27852(6)&-0.27966(3)& -0.279773(0)&0.8716\\ \hline
    & 7 & 42    &413,442	&229		& -0.280217 &-0.279841(0)   &-0.28021(0)&-0.27370(9)&-0.27701(4)&-0.27857(4)& -0.279841(0)&0.9631\\ \hline
    & 8 & 49    &5,557,206	&1,985		& -0.280580 &-0.279987(0)   &-0.28055(0)&-0.27498(0)&-0.27792(2)&-0.27927(7)& -0.279987(0)&0.9097\\ \hline
    & 9 & 56    &77,182,439	&17,487		& -0.280911 &-0.280117(0)   &-0.28086(1)&-0.27573(6)&-0.27853(3)&-0.27973(1)& -0.280117(0)&0.8348\\ \hline
    &10 & 63    &1,099,923,868	&178,665	& -0.281048	    &-0.280204(0)   &-0.28098(2)&-0.27593(3)&-0.27868(0)&-0.27985(8)&  -0.280204(0)&0.8119\\ \hline \hline 
    
1/9 & 4	& 27	&519		&5		& -0.248519 &-0.248141(0)   &-0.248516(0)&-0.24448(8)&-0.24613(3)&-0.24703(5)&-0.24768(0)&0.9736\\ \hline
    & 5 & 36    &7,483		&10		& -0.247875 &-0.247832(0)   &-0.247872(0)&-0.24312(7)&-0.24498(0)&-0.24616(3)&-0.24690(0)&0.9970\\ \hline
    & 6 & 45    &118,765	&91		& -0.249628 &-0.248674(0)   &-0.249595(0)&-0.24654(5)&-0.247916(6)&-0.248663(3)&-0.24908(7)&0.8277\\ \hline
    & 7 & 54    &1,999,265	&624		& -0.248967 &-0.248666(0)   &-0.248950(0)&-0.24481(6)&-0.24647(7)&-0.24758(6)&-0.24821(8)&0.9583\\ \hline
    & 8 & 63    &35,154,340	&7,105		& -0.249424
	    &-0.248855(0)   &-0.249381(0)&-0.24598(5)&-0.24746(2)&-0.24832(1)&-0.24872(5)& 0.8789\\ \hline 
    & 9 & 72    &638,724,335	&84,470		&-0.249817	    &-0.249018(0)  &-0.249752(1)&$-$&$-$&$-$&$-$&0.7762\\ \hline \hline
    
2/13& 4	& 18	&150		&3		& -0.288781 &-0.288751(0)   &-0.288751(1)&-0.28119(6)&-0.28530(6)&-0.28722(0)& -& 0.9992\\ \hline
    & 6 & 31    &17,002		&34		& -0.289873 &-0.289540(0)   &-0.289853(1)&-0.28394(6)&-0.28748(6)&-0.28881(9)& -0.28892(7)& 0.9642\\ \hline
    & 8 & 44    &2,502,617	&1,137		& -0.289885 &-0.289723(0)   &-0.289751(1)&-0.28344(3)&-0.28702(9)&-0.28852(9)& -0.28896(5)& 0.9820\\ \hline
    &10 & 57    &421,777,505	&85,250		& -0.290440	    &-0.290063(0)   &-0.290321(1)&-0.28449(3)&-0.28792(8)&-0.28922(3)& -0.28928(9)&$-$\\ \hline    
\end{tabular}}
\caption{\label{tab:energies_exact_liquids_crystals_small_N} Background-subtracted density-corrected per-particle lowest-Landau-level (LLL) Coulomb energies for various liquid and crystal states at filling factors $\nu=1/7$, $1/9$ and $2/13$. The fourth and fifth columns give the dimensions of the $L_{z}=0$ and $L=0$ subspaces respectively. The last column shows the absolute value of the overlap of the exact state with the trial Laughlin(L)/Jain(J) state\cite{Laughlin83,Jain89}. (The overlap with the exact state for the $\nu=2/13$ Jain state with $N=10$ particles is missing because we have not been able to obtain the Fock space decomposition of this Jain state due to certain technical difficulties. Given the close agreement between the energies, we expect the overlap to be large.) Some of the overlaps for 1/7 with the Laughlin state were previously quoted in Ref.~\cite{Shi19}. The numbers in the parenthesis are the statistical uncertainty of the Monte Carlo calculation of the energies. The $^{6}$CF number at $\nu=1/7$ are the same as that of the Laughlin state. The symbol ``$-$" indicates numbers which are currently unavailable.}
\end{table*}

\section{DMRG studies}
\label{sec: DMRG}
\subsection{DMRG method}

Since the development of the DMRG algorithm~\cite{White92, Schollwock05} and its description in terms of matrix product state (MPS) representation~\cite{Schollwock11}, this algorithm has become a reliable tool to precisely calculate the ground-state correlation functions and energies for one-dimensional quantum many-body systems. Through mapping the two-dimensional electrons under a strong magnetic field in the lowest Landau level to a one-dimensional lattice model, DMRG algorithms have been used to investigate the FQHE liquids and composite fermion Fermi sea~\cite{Shibata01, Feiguin08, Zaletel13, Zaletel15, Geraedts16}. Here we apply the finite DMRG approach to the cylindrical geometry using the Landau gauge, following Zaletel, Mong, and Pollmann~\cite{Zaletel13}. 

{A remark is in order on why we use the cylindrical geometry rather than the spherical geometry for our DMRG study. In comparison to the cylinder, on the sphere, it is not possible to go to very large systems using finite DMRG. While we can go to 25 particles at $\nu=1/3$ using DMRG (also see Ref.~\cite{Zhao11}), the largest system accessible to us on the sphere at filling factor 1/7 is 12 particles. The main reason for the difference in the system sizes accessible to the two geometries owes to their single-particle wave functions~\cite{Hu12}. On the cylinder,  single-particle wave functions are Gaussian wave packets, which are highly localized. As a result, the coefficient amplitudes of the electron-electron interaction drop rapidly, allowing us to truncate lots of terms of the Hamiltonian for DMRG and get reasonable results of large sizes. For the sphere, on the other hand, the single-particle wave functions are not highly localized, requiring more states to be retained for achieving comparable accuracy.}

{The cylindrical geometry is topologically equivalent to a parallelogram in the complex plane with periodic boundary conditions in one direction and open boundary conditions in the other. For a finite-size system with $N$ particles at filling $\nu = 1/m$, the number of orbitals (i.e., single-particle eigenstates) is given by $N_{\rm orb} =m(N-1)+1$, which is the same as that on the spherical geometry~\cite{Ortiz13}. Once the particle number and filling factor are fixed, the geometry of the cylinder is specified by the circumference $L$ in the periodic direction. ($L$ has also been used to denote the total orbital angular momentum in the spherical geometry; we hope that its meaning is clear by context.) In the open boundary direction, the particles would distribute in an approximate range of $[-\pi N_{\rm orb}\ell^2/L,\pi N_{\rm orb}\ell^2/L]$, out of which the density would decay as an exponential function. Rather than using the usual Cartesian coordinates $(x,y)$, we choose a ``reduced coordinate" system $(\theta_1,\theta_2)$, which is defined as: $x+iy=\theta_1L+\theta_2L\tau$. Here, $\tau=\tau_1+i\tau_2$ is a parameter in analogy to the modular parameter~\cite{Gunning62} in torus. 

As shown below, the Hamiltonian only depends on $N_{\rm orb}$ and $L$, so the choice of $\tau$ for the cylinder is just a matter of parametrization of the particle coordinates. Within the reduced coordinate system, we choose a hexagonal cylinder defined by $\tau=e^{i\pi/3}$ and $L=\sqrt{2\pi N_{\rm orb}\ell^2/\tau_2}$. The advantage of this choice is that it allows a convenient visualization of the triangular crystal. Since $\theta_1$ is along the circumference of the cylinder, it lies between $\left[ -0.5,0.5\right)$. The coordinate $\theta_2$ may have some extension outside of $\left[ -0.5,0.5\right)$, but there the density of particles is small and decays very rapidly to zero. Using the Landau gauge $A=(-y, 0) / \ell^{2}$, we can obtain the single particle wave-function in the LLL~\cite{Zaletel13}: 
\begin{align}
\varphi_{n}(\theta_{1},\theta_{2}) =\left(  2\tau_{2}
N_{\rm orb}\right)  ^{1/4}e^{i2\pi n\left(  \theta_{1}+\theta_{2}\tau_{1}\right)
}e^{-\pi\tau_{2}N_{\rm orb}\left(  \theta_{2}-\frac{n}{N_{\rm orb}}\right)  ^{2}}
\end{align}
where $N_{\rm orb}$ is the number of orbits and  index $n$ takes values $0,1,\cdots,N_{\rm orb}-1$. The wave function $\varphi_{n}(\theta_{1},\theta_{2})$ satisfies the periodic boundary condition $\varphi_{n}(\theta_{1}+1,\theta_{2})=\varphi_{n}(\theta_{1},\theta_{2})$, which can be viewed as a one-dimensional lattice model with lattice constant $1$.

As noted in Ref.~\cite{Johri15}, the long range nature of the Coulomb interaction leads to divergences in the  cylindrical geometry. These can be avoided by choosing the interaction $V( \mathbf{r})=\frac{e^{-\mu r}}{r}$, which reduces to the Coulomb interaction for $\mu=0$. [All energies in this paper are quoted in units of  $e^{2} / (\epsilon \ell)$.] The total energy is composed of three terms:
\be
H_{\rm total}=H_{\rm ee}+H_{\rm eb}+H_{\rm bb}
\ee
The last term on the right-hand side, which represents the background-background interaction, is constant and can be omitted in the calculation of density profiles and pair-correlation functions. The first two terms represent the electron-electron interaction and electron-background interaction. In the second quantized form the various terms in the Hamiltonian can be expressed as: 
\begin{equation}
H_{ee}={\sum\limits_{n,k\geq\left\vert m\right\vert }^{N_{\rm orb}}}V_{mk}^{ee}c_{n+m}^{\dagger}c_{n+k}^{\dagger}c_{n+m+k}c_{n}
\end{equation}
\be
\label{Hamiltonian1}
V_{mk}^{ee}=2(V_{mk}-V_{km})
\ee
\begin{equation}
V_{mk}=\frac{e^{-{2\pi^2m^2\ell^2\over L^2}}}{L}\int_{0}^{\infty}d q_{2}\frac{e^{-\frac{q_{2}^{2}}{2}}\cos\left(  2\pi k q_{2}/L\right)  }{\sqrt{\left(  \frac{2\pi m}{L}\right)  ^{2}+q_{2}^{2}+\mu^{2}}}
\end{equation}
\begin{equation}
\label{Hamiltonian2}
H_{eb}=\sum_{n}V_{n}^{eb}c_{n}^{\dagger}c_{n}
\end{equation}
\begin{align}
V_{n}^{eb}=-{2N\over N_{\rm orb}L}\sum_{j=0}^{N_{\rm orb}-1}\int_{0}^{\infty}dq_{2} {e^{-q_2^2/2}\cos{2\pi q_2(j-n)\over L}\over\sqrt{q_2^2+\mu^2}}\label{eq:Veb}
\end{align}
\begin{equation}
\label{Hamiltonian3}
H_{bb}=\sum_{n}V^{bb}c_{n}^{\dagger}c_{n}
\end{equation}
\begin{align}
V^{bb}={N\over N_{\rm orb}^2L}\sum_{j,n=0}^{N_{\rm orb}-1}\int_{0}^{\infty}dq_{2} {e^{-q_2^2/2}\cos{2\pi q_2(j-n)\over L}\over\sqrt{q_2^2+\mu^2}}
\end{align}
where $c_n^\dagger$ and $c_n$ are the creation and annihilation operators for single particle orbital $\varphi_n$. Because the electron charge distribution of the LLL for a finite system is nonuniform, we choose the background charge distribution to be proportional to the electron charge distribution of lowest filled Landau level, i.e. $\rho_{\rm b}(\mathbf{r}) =\frac{N}{N_{\rm orb}}\sum_{n=0}^{N_{\rm orb}-1}\left\vert \varphi_{n}\left(\mathbf{r}\right)  \right\vert ^{2}$. We note that the Hamiltonian has no dependence on $\tau$, which is natural because how we parametrize the coordinates on cylinder is irrelevant to any physical observable.

The pair-correlation function is a natural quantity that can distinguish between the Wigner crystal and the FQH liquid states. The pair-correlation function for our finite-size system is defined in Eq.~(\ref{eq:gr}) where we take $\rho_0={N\over 2\pi N_{\rm orb}\ell^2}$ as the mean density. 

We note that in the above Hamiltonian we have not included the self-interaction energy, which is the interaction energy between an electron and its images in the periodically repeated unit cells~\cite{Yoshioka83}. The self-interaction energy for torus is a negative constant, independent of the wave function, and does not affect the pair-correlation function. Because of the presence of edges in the cylindrical geometry, the self-interaction energy has a slight dependence on the electron position, and hence the wave function. However, compared to the bare electron background in Eq.~(\ref{eq:Veb}) the corrections are proportional to $1/N$, which we have neglected. We have explicitly checked for $N=6$, 9, and 12 for $\nu=1/7$ that inclusion of the self-interaction energy has no discernible effect on the pair-correlation function. 

As another note, the cylindrical geometry is not useful for determining the ground state energy. The primary reason is that corrections due to the presence of two edges are very strong, as evident by the presence of a charge density wave spanning the entire length of the cylinder.  A proper thermodynamic limit for the energy would require going to large enough systems where the density in the interior becomes uniform, which is not the case for the systems that we have studied. This issue can be avoided and accurate Coulomb energies can be obtained by using {\it infinite} DMRG (iDMRG), as shown by Zaletel {\it et al.}~\cite{Zaletel15}.

The DMRG algorithm is based on the ITENSOR library ~\cite{ITensor19}.

\subsection{DMRG results}

\begin{figure*}[tbh]
\centering
\includegraphics[width=2.1\columnwidth]{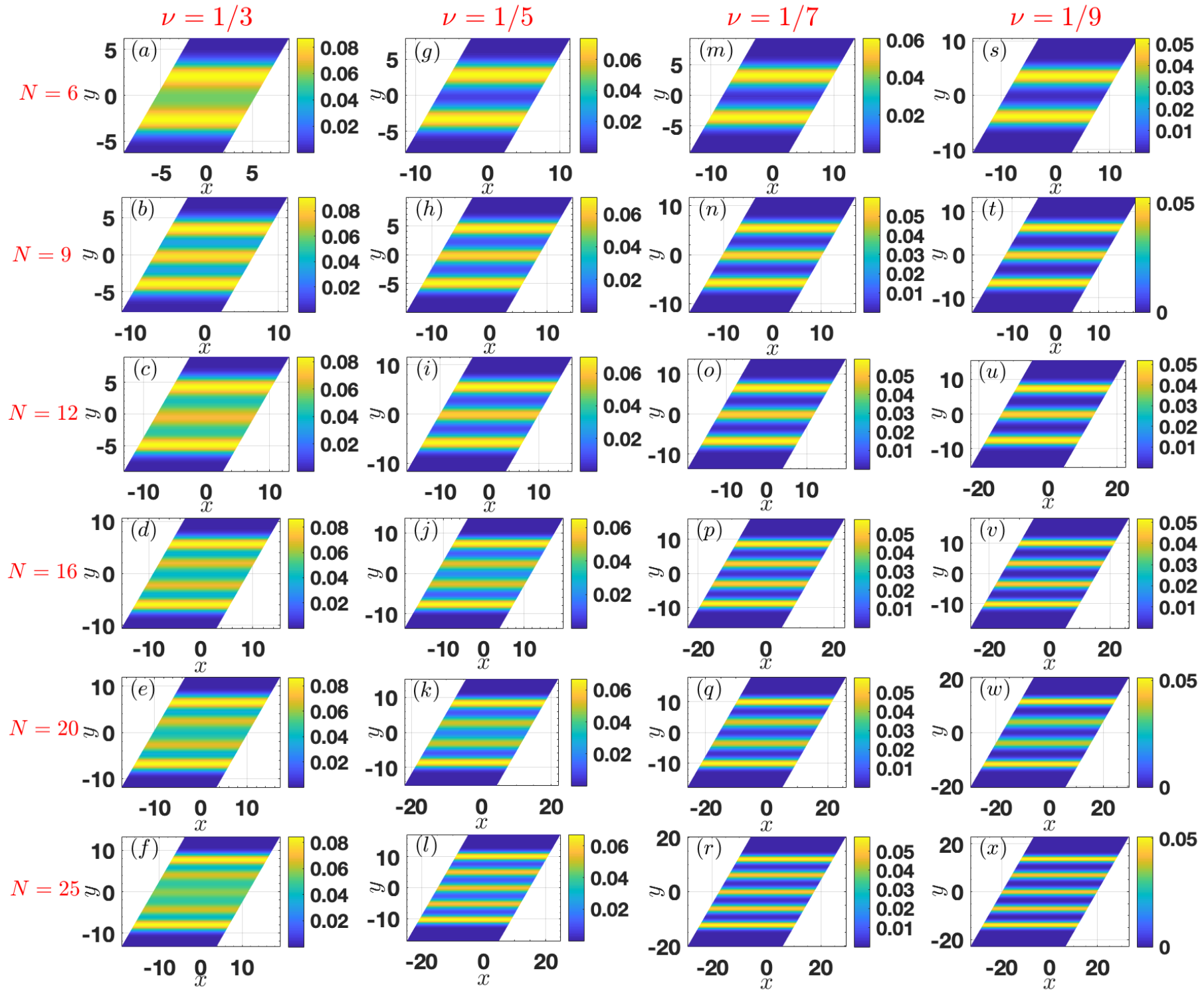}
\caption{The electron density profiles on a hexagonal cylinder. The columns show results for filling factors $\nu=1/3$, 1/5, 1/7, and $1/9$. The rows correspond to $N=6$, 9, 12, 16, 20 and 25 particles. We take periodic boundary condition for the horizontal direction and open boundary condition for the vertical direction. We choose $L=\sqrt{4\pi N_{\rm orb}\ell^2/\sqrt{3}}$, $\tau=e^{i\pi/3}$ and $\mu=0.0001$.}
\label{NewFigDensity}
\end{figure*}

\begin{figure*}[tbh]
\centering
\includegraphics[width=2.1\columnwidth]{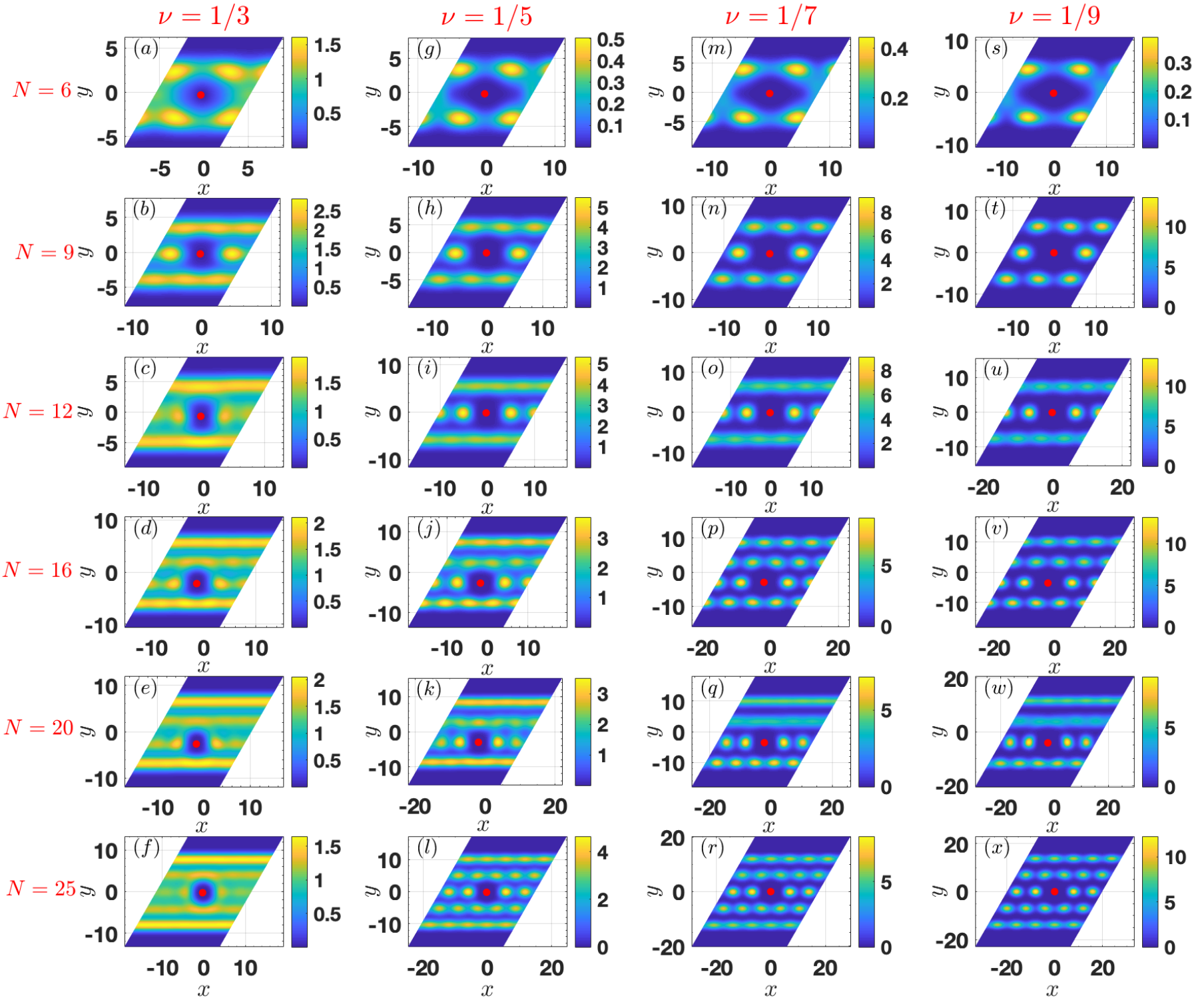}
\caption{The pair-correlation functions on a hexagonal cylinder. The columns show results for filling factors $\nu=1/3$, 1/5, 1/7, and $1/9$. The rows correspond to $N=6$, 9, 12, 16, 20 and 25 particles. The red dot indicates the position of the fixed electron. We take periodic boundary condition for the horizontal direction and open boundary condition for the vertical direction. We choose $L=\sqrt{4\pi N_{\rm orb}\ell^2/\sqrt{3}}$, $\tau=e^{i\pi/3}$ and $\mu=0.0001$.
}
\label{NewFigPCF}
\end{figure*}

Figures~\ref{NewFigDensity} and \ref{NewFigPCF} show the density profiles and pair-correlation functions for $\nu=1/3$, 1/5, 1/7 and 1/9 for $N=6$, 9, 12, 16, 20 and 25 particles. The bond dimension is kept at least $\chi=6000$ in the truncation of bases, which allows the singular value decomposition truncation error to be smaller than $10^{-7}$. We choose $\mu=0.0001$; we have found that our results are insensitive to the value of $\mu$ so long as $\mu\leq 0.01$, and are thus represent the behavior for the Coulomb interaction.  
We have also confirmed that the pair-correlations and orbital occupations of ground states through DMRG agree perfectly with those obtained from exact diagonalization studies, wherever the latter results are available. 
In the calculation of the pair-correlation function, the position of the fixed electron is taken on one of the interior maxima in density (except for $N=6$), so that it would coincide with the position of a crystal site, should a crystal be stabilized. (This maximizes the chances for a crystal.) In the background information~\cite{DMRGmovie}, we show how the pair-correlation function $g(\vec{r},\vec{r}')$ changes as we vary the position of $\vec{r}$ (indicated by a red dot in the figures) along the length of the cylinder.

The question, of course, is whether the system is a crystal or a liquid in the thermodynamic limit. We note that the density oscillations along the length of the cylinder are dictated by the open boundary conditions; of relevance here are the correlations in the horizontal, i.e. periodic, direction of the cylinder. The $\nu=1/3$ state looks like a liquid for all particle numbers, so here there is no ambiguity regarding its liquid nature. For $\nu=1/5$, 1/7, and 1/9, the pair-correlation function indicates at least short-range crystalline correlations, which strengthen with lowering $\nu$. Furthermore, the crystalline structure depends on the particle number: systems with $N=9$, 16, and 25 particles look more crystalline than those with 6, 12, and 20 particles. The reason is that $N=9$, 16, and 25 particles allow for the possibility of a triangular crystal for our geometry. We can make the following observations. First, for $N=6$, 12, and 20, the system looks like a liquid; for example, for $N=20$, very weak crystalline correlations are seen in the region of high density near the top of the cylinder.  Furthermore, even for $N=9$, 16, and 25 particles, for both $\nu=1/7$ and $\nu=1/9$ the crystalline correlations decay as we go to larger systems. For example, the top row for $N=25$ is more liquid-like than the top row for $N=16$ particles.

Our study in the cylindrical geometry thus again indicates that the competition between FQHE liquid and the crystal phase at low filling is rather subtle. It does clarify, unambiguously, that the state has strong short-range crystalline correlations, but does not rule out a liquid state in the thermodynamic limit.

\section{Exact diagonalization studies}
\label{subsec: EDresults}

We have performed calculations for larger systems than before, going to $N=10$ for $\nu=1/7$, and to $N=9$ for $\nu=1/9$ and $1/11$. These systems correspond to Hilbert space dimensions of $1.1\times10^9$, $0.6\times10^9$, and $3.4\times10^9$, respectively. (To calculate the gaps to charged excitations we need to increase the flux by one unit above the ground states, making the size of the diagonalized matrices even  larger.)
 
The key technical innovation used to efficiently diagonalize such large matrices is that in the on-the-fly matrix-times-vector multiplication done at each Lanczos iteration we exploit the following symmetry of the two-body interaction: states with the same $N-2$ occupied orbitals and different only by the pair of states occupied by the last two electrons form a sequence distinguished by a single two-body scattering. Importantly, a two-body scattering connects any two states of this sequence. Hence, there is a nonzero matrix element for each element of the corresponding submatrix, and it is equal to the corresponding two-body scattering amplitude. Since the most time-consuming part of the matrix-times-vector product is finding the location (row and column) of a given amplitude in the matrix, filling the matrix by consecutive submatrices is a relatively fast approach. On the downside, efficient parallelization with this method requires more memory than in approaches which divide the matrix into disconnected rows. For the systems with dimensions exceeding $10^9$, such computation with 24-fold parallelization requires at least 512 GB of RAM (widely accessible today) and yields speeds of one Lanczos iteration per couple of hours (real time) and a total time of a couple of weeks to typically reach convergence for the ground state (the convergence criterion is that the energy difference between successive Lanczos iterations is less than $10^{-9}$). 

We find that the exact Coulomb ground states at $1/7$, $1/9$, $1/11$, and $2/13$ have $L=0$ for all $N$ studied. This is significant because a ground state with $L\neq 0$ would indicate a compressible state (e.g. a crystal). The energies of finite-size systems and their overlaps with the model Laughlin or Jain states are quoted in Table~\ref{tab:energies_exact_liquids_crystals_small_N}. The exact Coulomb ground states have significant overlap with the Laughlin or Jain wave functions at these special fillings. In Fig.~\ref{fig:pair_correlations} we compare the pair-correlation function $g(r)$~\cite{Jain07} of the exact LLL Coulomb ground state with that of the Laughlin state. We find that the $g(r)$ for these two states are in reasonable agreement with each other. Possibly there are crystalline correlations at short distances, with the strength of these correlations being much stronger in the exact LLL Coulomb ground state than in the Laughlin state.

\begin{figure}[tbh]
\begin{center}
\includegraphics[width=0.47\textwidth,height=0.23\textwidth]{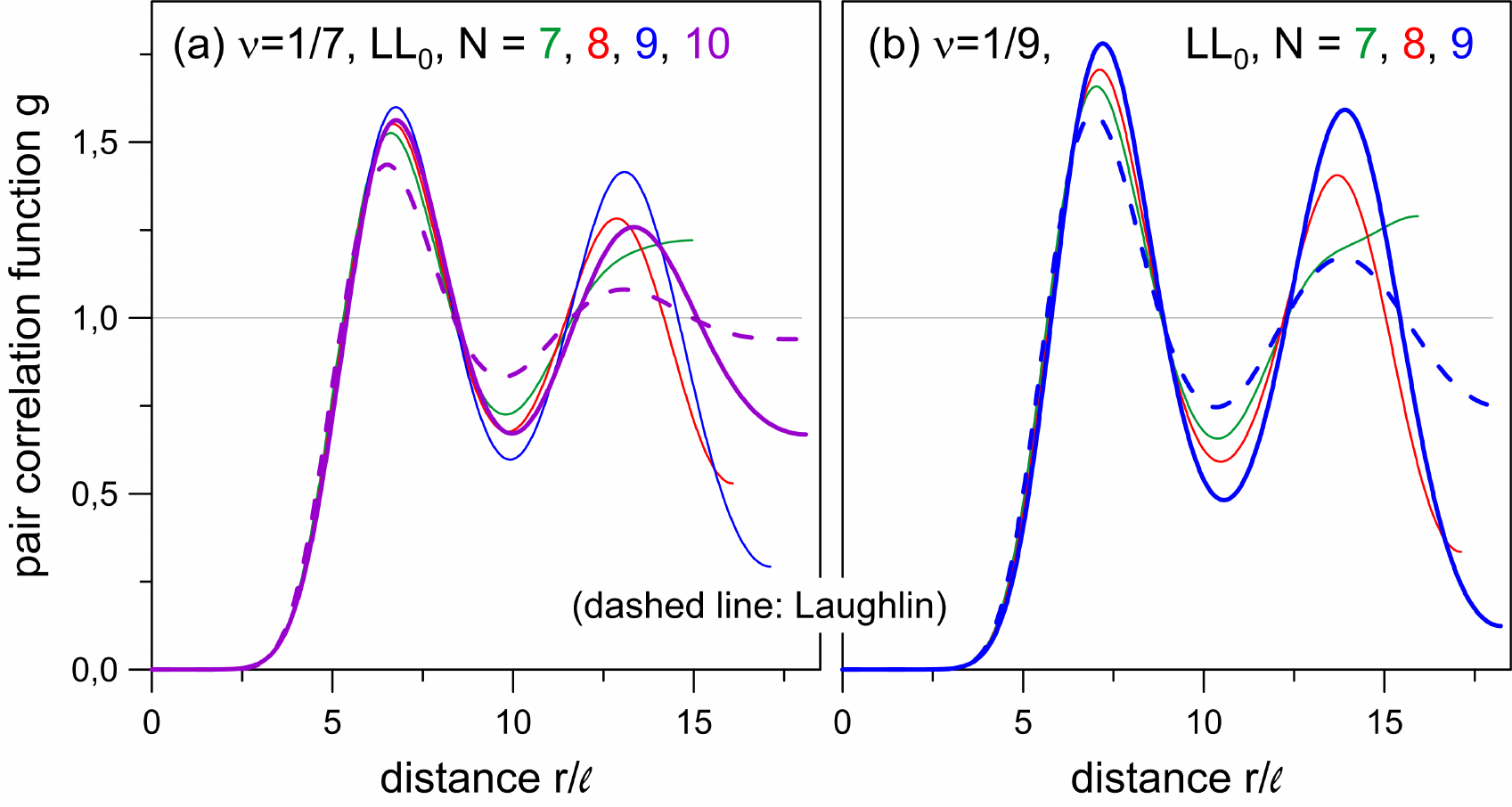} 
\caption{(color online) Pair-correlation function $g(r)$ for the exact lowest-Landau-level Coulomb ground state (thick line) and Laughlin state (dashed line) at $\nu=1/7$ (left panel) and $1/9$ (right panel).} 
\label{fig:pair_correlations}
\end{center}
\end{figure}

The extrapolation of the exact LLL Coulomb ground state energies at fillings $\nu=1/3,~1/5,~1/7$, and $1/9$ are shown in Fig.~\ref{fig:extrapolations_ground_state_energies}. We multiply the per-particle energies by a factor of $\sqrt{2Q\nu/N}$ before extrapolating to $N\rightarrow \infty$~\cite{Morf86}. This factor corrects for the finite-size deviation of the electron density from its thermodynamic value, thus providing a more accurate extrapolation.

\begin{figure}[tbh]
\begin{center}
\includegraphics[width=0.47\textwidth,height=0.23\textwidth]{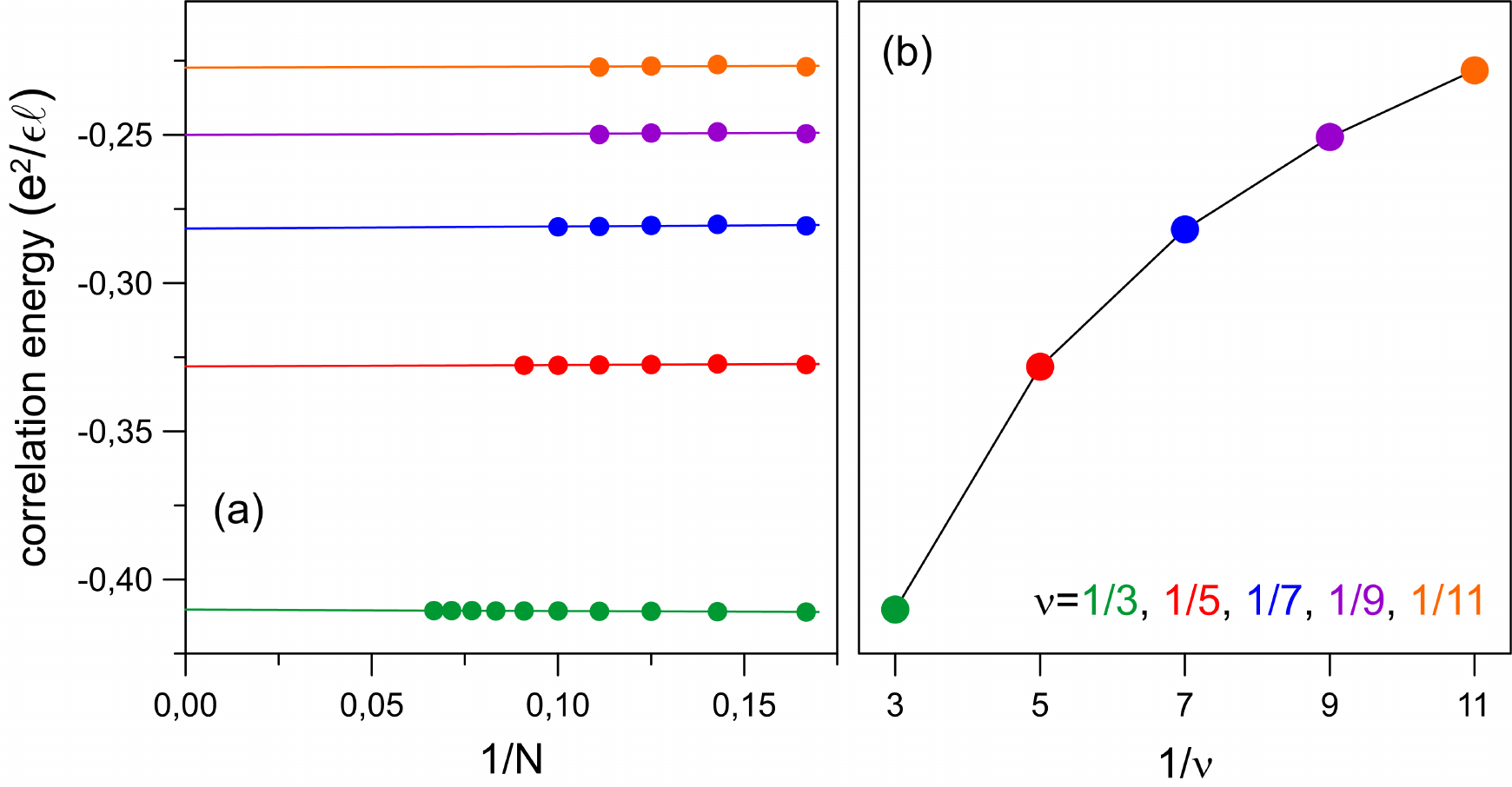} 
\caption{(color online) Thermodynamic extrapolation of the exact lowest-Landau-level Coulomb ground-state energies for fillings $\nu=1/3$, 1/5, 1/7, 1/9, and 1/11. Left panel (a) shows the extrapolation obtained from energies of finite-size systems in the spherical geometry and the right panel (b) shows the thermodynamic energies as a function of $1/\nu$. The correlation energies include interaction with the positively charged background and have been density corrected. All energies are quoted in units of $e^2/(\epsilon\ell)$. } 
\label{fig:extrapolations_ground_state_energies}
\end{center}
\end{figure}

An important characteristic of an incompressible state is that it costs a finite energy to create charged excitations. Figure~\ref{fig:extrapolations_charge_gap_Laughlin_states} depicts the charge gap as a function of $1/N$ at fractions of the type $1/(2p+1)$. The charge gap at these Laughlin fractions is defined as the sum of the energies of a quasihole (QH) and a quasiparticle (QP), which in turn are determined from exact diagonalization at $2Q=(2p+1)(N-1)\pm 1$, respectively. We find that the total orbital angular momentum quantum number of the exact ground state at $2Q=(2p+1)(N-1)\pm 1$ is consistent with that predicted by the CF theory for all the systems considered in this work. The charge gap is equal to the energy required to create a far separated pair of quasihole and quasiparticle. From the extrapolation, we obtain the charge gap in the thermodynamic limit. While there are some finite-size fluctuations, at least from the system sizes available to us, the charge gaps appear to extrapolate to non-zero values for both $\nu=1/7$ and $\nu=1/9$. The extrapolated thermodynamic values of the charge gap are shown in Fig.~\ref{fig:extrapolated_charge_gap} for several Laughlin fractions.

\begin{figure}[tbh]
\begin{center}
\includegraphics[width=0.47\textwidth]{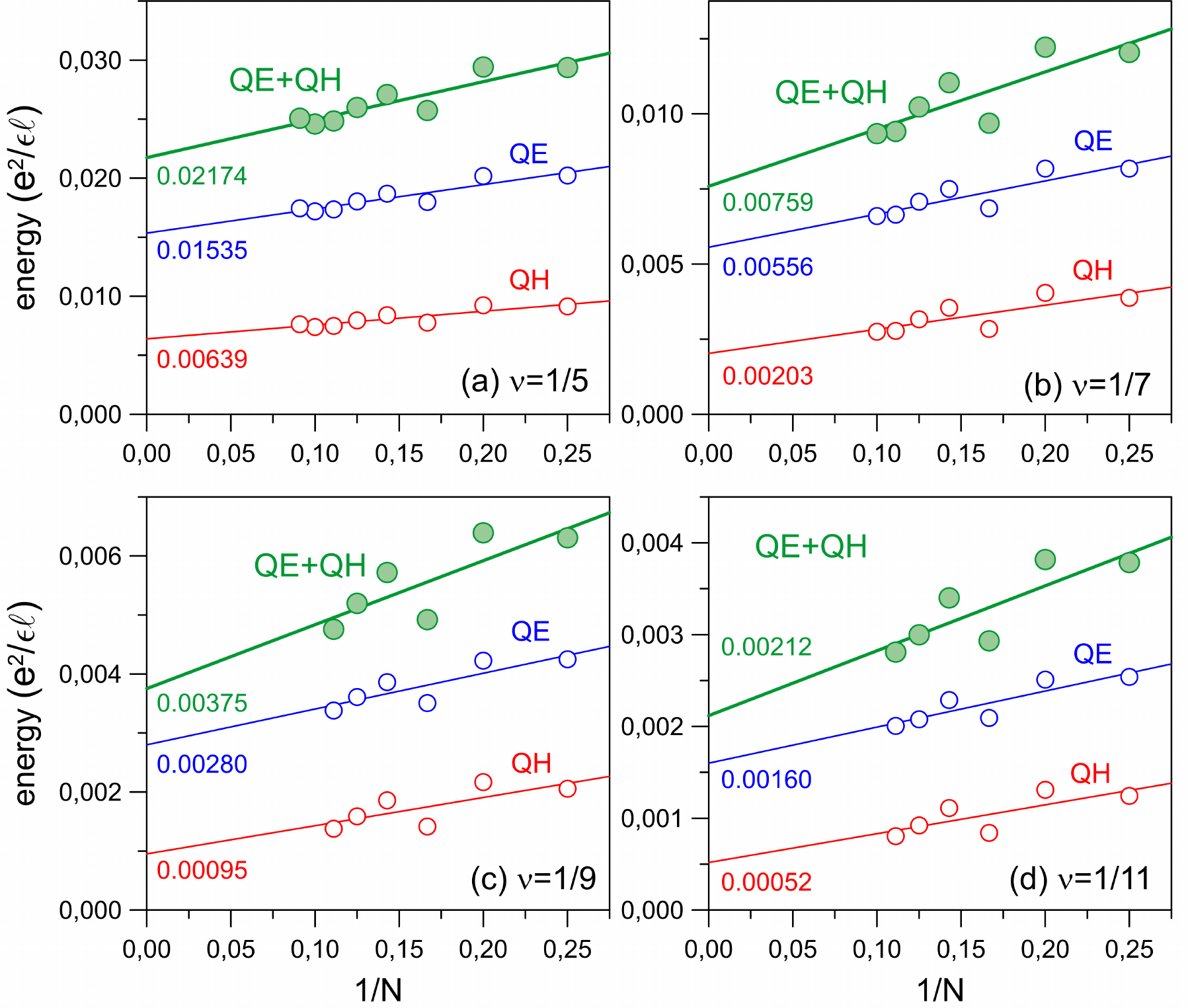} 
\caption{(color online) The energy of the quasihole (QH), quasiparticle (QP), and their sum for the lowest-Landau-level Coulomb state at Laughlin fractions as a function of $1/N$, where $N$ is the number of electrons. All the energies quoted in units of $e^2/(\epsilon\ell)$.} 
\label{fig:extrapolations_charge_gap_Laughlin_states}
\end{center}
\end{figure}

\begin{figure}[tbh]
\begin{center}
\includegraphics[width=0.47\textwidth,height=0.41\textwidth]{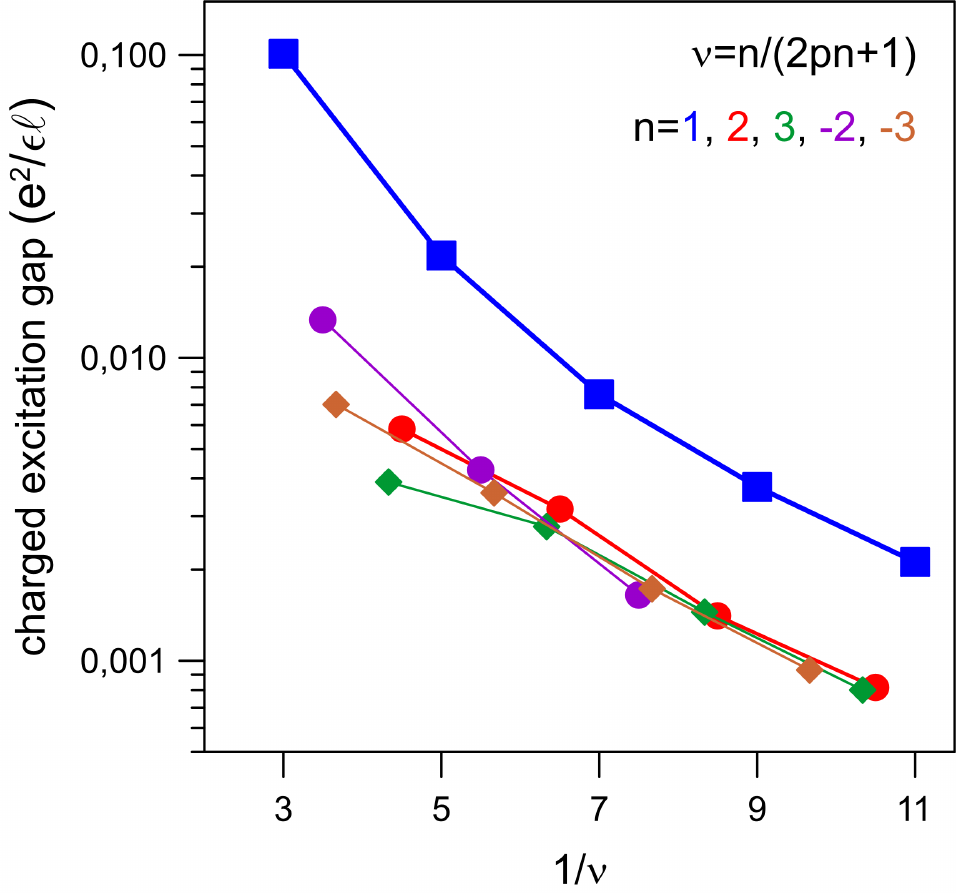} 
\caption{(color online) Exact lowest-Landau-level Coulomb charge gap, which is the energy required to create a far-separated quasiparticle-quasihole pair, for fillings $\nu=n/(2pn+1)$. The extrapolated gaps were obtained from a linear fit in $1/N$ of gaps of finite-size systems in the spherical geometry (see Fig.~\ref{fig:extrapolations_charge_gap_Laughlin_states} for extrapolations of the charge gap for Laughlin fractions). All energies are quoted in units of $e^2/(\epsilon\ell)$. The charge gap for $1/3$ has been reproduced from Ref.~\cite{Balram20}.} 
\label{fig:extrapolated_charge_gap}
\end{center}
\end{figure}

Next, we turn to neutral excitations at filling factors $1/(2p+1)$. The CF theory predicts that the lowest-energy neutral excitations are excitons of composite fermions, consisting of a single CF-particle-hole excitation. (The neutral exciton mode has also been called the magnetoroton mode.) The CF theory predicts that the exciton branch extends from angular momentum $L=2$ to $L=N$. Our exact diagonalization study shows a clearly identifiable low-energy neutral mode extending from $L=2$ to $L=N$. The dispersions of the magnetorotons at $1/7$ and $1/9$ are shown in Fig.~\ref{fig:roton_dispersions}. Due to strong finite-size effects, we have not been able to obtain a reliable thermodynamic extrapolation of the smallest neutral gap at low fillings. 

\begin{figure}[tbh]
\begin{center}
\includegraphics[width=0.47\textwidth,height=0.23\textwidth]{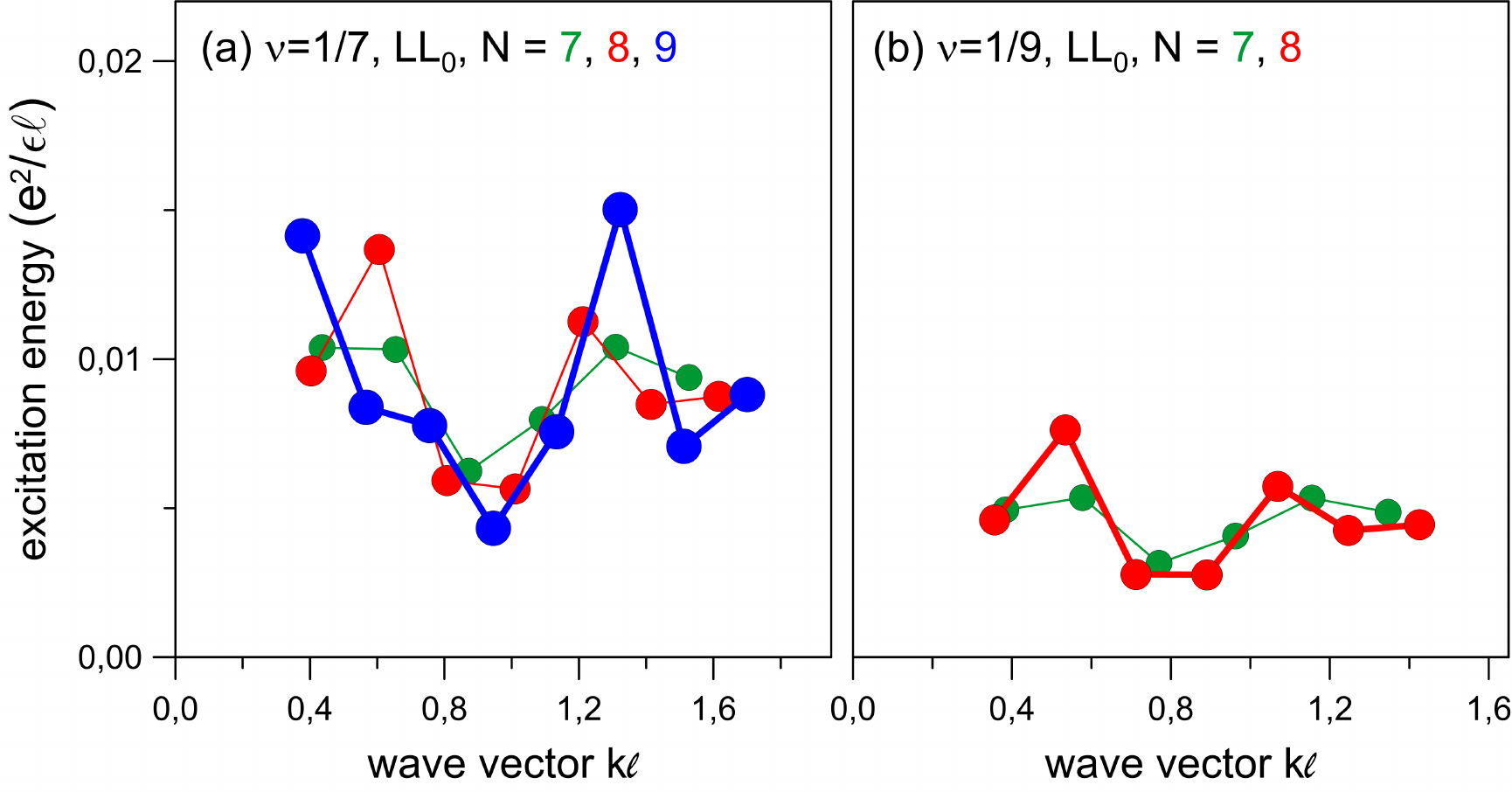} 
\caption{(color online) Magnetoroton dispersions of the exact lowest-Landau-level Coulomb state at $1/7$ (left panel) and $1/9$ (right panel) as a function of the dimensionless wave vector $k \ell=L \ell/R$, where $L$ is the total orbital angular momentum and $R$ is the radius of the sphere. All the energies quoted in units of $e^2/(\epsilon\ell)$ and are measured relative to the ground state energy.} 
\label{fig:roton_dispersions}
\end{center}
\end{figure}

Figure~\ref{fig:fix_N_sweep_2Q} shows a plot of Coulomb ground-state energy in the LLL for $N=10$ electrons as a function of the flux $2Q$. The downward cusps in the correlation energies obtained from exact diagonalization at $\nu=1/7$ and $\nu=2/13$ suggest that the states here are incompressible. These cusps do not have a natural explanation in terms of a crystal ground state.

\begin{figure}[tbh]
\begin{center}
\includegraphics[width=0.47\textwidth,height=0.23\textwidth]{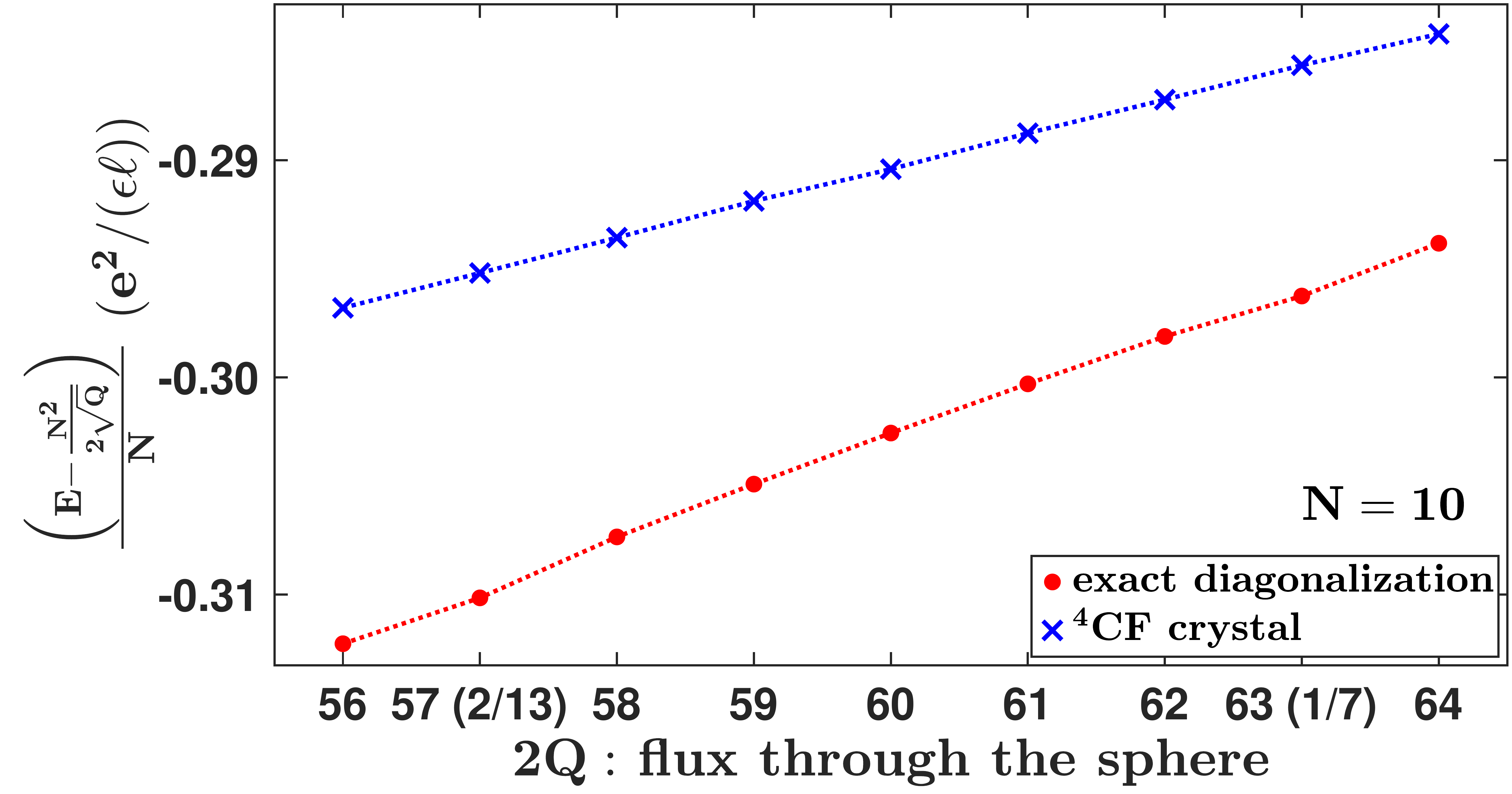} 
\caption{(color online) A plot of the lowest-Landau-level Coulomb energies for $N=10$ electrons as a function of the flux $2Q$ on the sphere. Cusps in the ground-state energies can be seen at the special filling factors $\nu=1/7$ and $2/13$ in the exact diagonalization data. This should be contrasted from the energies of the best crystal (that formed from composite fermions carrying four vortices), which is a smooth function of the filling factor. The dotted lines are a guide to the eye.} 
\label{fig:fix_N_sweep_2Q}
\end{center}
\end{figure}

Considering all these results we conclude that exact diagonalization studies on small systems are fully consistent with the formation of incompressible FQHE states at $\nu=1/7$ and $\nu=1/9$. 

\subsection{Entanglement spectra}
\label{subsec: entanglement_spectra}
We have also studied the edge excitations of the exact ground state by evaluating its entanglement spectrum. The entanglement spectrum has been useful in characterizing many FQH states because the counting of low-lying entanglement levels characterizes the topological order of the state~\cite{Li08}. To evaluate it from the ground state wave function $|\Psi\rangle$, we divide the system into two sub-systems $A$ and $B$, so that the state can be written as: $|\Psi\rangle=\sum_{i,j} c_{ij} |\Psi^{A}_{i}\rangle \otimes |\Psi^{B}_{j}\rangle$ where $|\Psi^{A}_{i}\rangle$ and $|\Psi^{B}_{j}\rangle$ are the basis states for the $A$ and $B$ sub-systems respectively and $c_{ij}$ are the expansion coefficients. We then do a Schmidt decomposition to obtain: $|\Psi\rangle=\sum_{k} e^{-\xi_{k}/2} |\Psi^{A}_{k}\rangle \otimes |\Psi^{B}_{k}\rangle$, where $\xi_{k}$ are the entanglement eigenvalues which form the entanglement spectrum. Figure~\ref{fig:entanglement_spectra_Laughlin_Coulomb_1_7} shows the orbital entanglement spectrum~\cite{Haque07} of the exact LLL Coulomb ground state at $\nu=1/7$ for a system of $N=8$ electrons at a flux of $2Q=49$ on the sphere (results for smaller systems are similar). The counting of the low-lying entanglement levels of the exact LLL Coulomb ground state $\nu=1/7$ is identical to that of the Laughlin state which indicates that the two states lie in the same topological phase.
%%%%%%%%%%%%%%%%%%%%%%%%%%%%%%%%%%%%%%%%%%%%%%%%%%%%%%%%%%%%%%%%%%%%%%%%%
\begin{figure}[tbh]
\begin{center}
\includegraphics[width=0.47\textwidth]{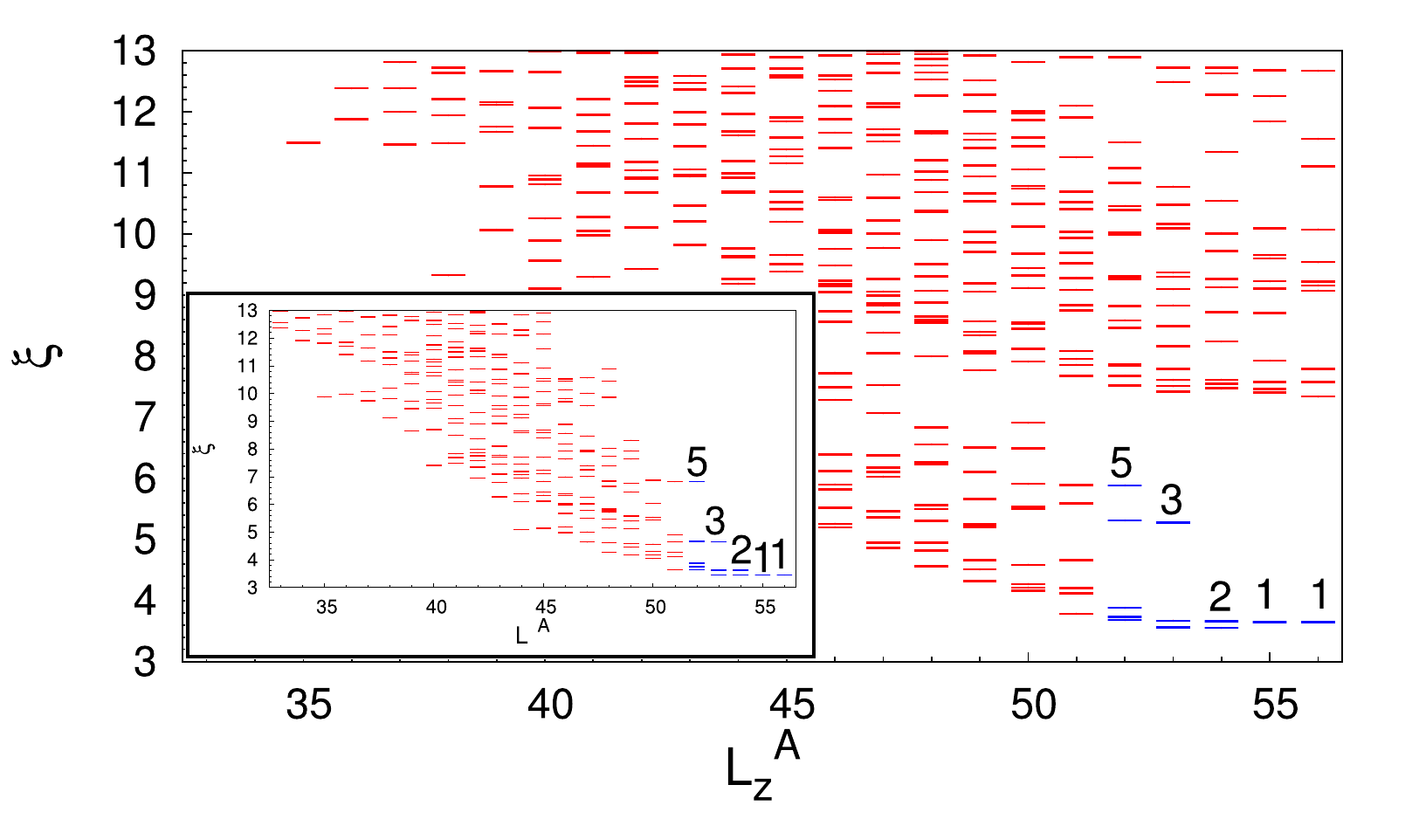} 
\caption{(color online) Orbital entanglement spectrum at $\nu=1/7$ of the exact lowest-Landau-level Coulomb ground state and the Laughlin state (inset) for $N=8$ electrons at a flux $2Q=49$ on the sphere. The two subsystems $A$ and $B$ with respect to which the entanglement spectrum is calculated have $N_{A}=N_{B}=4$ electrons and $l_{A}=l_{B}=25$ orbitals. The entanglement levels are labeled by the $z$-component of the total orbital angular momentum of the  subsystem $A$, $L_{z}^{A}$. The counting of low-lying levels (from $L_{z}^{A}=56$ and going from from right to left) goes as $1,1,2,3,5,\cdots$, which matches with the $U(1)$ chiral boson counting~\cite{Regnault15}.}
\label{fig:entanglement_spectra_Laughlin_Coulomb_1_7}
\end{center}
\end{figure}

\section{Discussion}
\label{sec: discussion}

In this paper, we have revisited the issue of the nature of the state, i.e., whether it is an incompressible liquid or a crystal, at small filling factors. Previous calculations based on either variational wave functions or exact diagonalization on small systems had suggested a crystal phase for $\nu\lesssim 1/6$.  We find that the issue is more delicate than previously thought. A variational study finds that the energies of the CF crystal and a renormalized FQH liquid are too close to call. DMRG studies in cylindrical geometry show results that fluctuate with the particle number and do not rule out a liquid state in the thermodynamic limit. 

Most informative, we believe, are exact diagonalization studies in the spherical geometry, which are fully consistent with incompressible FQHE states at these fractions. The quantum numbers of the exact ground states, their neutral excitations, their quasiparticles, and quasiholes, are precisely what is expected from the FQHE theory; their gaps extrapolate to nonzero values in the thermodynamic limit, and, most importantly, these states have significant overlaps with the Laughlin / Jain wave functions. One might argue that the spherical geometry is not the most friendly to a crystal, as a triangular crystal cannot be wrapped on the surface of a sphere without causing defects, which might disfavor a crystal. However, while one can see why a crystal phase would get distorted on a sphere, we do not see any reason why it would turn into a normal FQHE liquid.  Our systems are large enough that there is ample freedom for obtaining a state that is not uniform ($L\neq 0$) or a state that is not an FQHE liquid. The fact that the actual ground states and their excitations are so well described by known FQHE physics is thus nontrivial. 

Based on these considerations, we conclude that the states at $\nu=1/7$, 2/13, and 1/9 are likely to be FQHE liquids in the thermodynamic limit. Our calculations suggest that rather than forming a full-fledged crystal immediately below $\nu\lesssim 1/6$, the system finds it advantageous to form, at certain filling factors, the FQHE liquid phase with strong short-range crystalline correlations. At what filling factor we get to a true crystal phase is beyond the scope of this paper.

Our work suggests the following scenario. In the absence of disorder, the ground states at filling factors of the type $\nu=n/(6n\pm 1)$ and $\nu=n/(8n\pm 1)$ are incompressible FQH liquids, but slightly away from these filling factors the crystal prevails (analogously to what happens near $\nu=1/5$). As a result, the filling factor variation due to disorder creates domains of crystal. For small disorder the liquid will percolate, producing standard FQHE.  For disorder greater than some filling-factor-dependent critical value, the variation in the filling factor is sufficiently large to prevent percolation of the FQH liquid. In this case, an insulator with exponentially large resistance will occur at low temperatures. Nonetheless, the presence of FQH domains produces, at the special filling factors, resistance minima, because the FQH domains provide edge states that enable dissipationless transport along parts of the sample. This scenario is consistent with the experimental observations. A schematic phase diagram suggested by our work is depicted in Fig.~\ref{phase-diagram}.

\begin{figure}[t]
\begin{center}
\includegraphics[width=0.48\textwidth,height=0.28\textwidth]{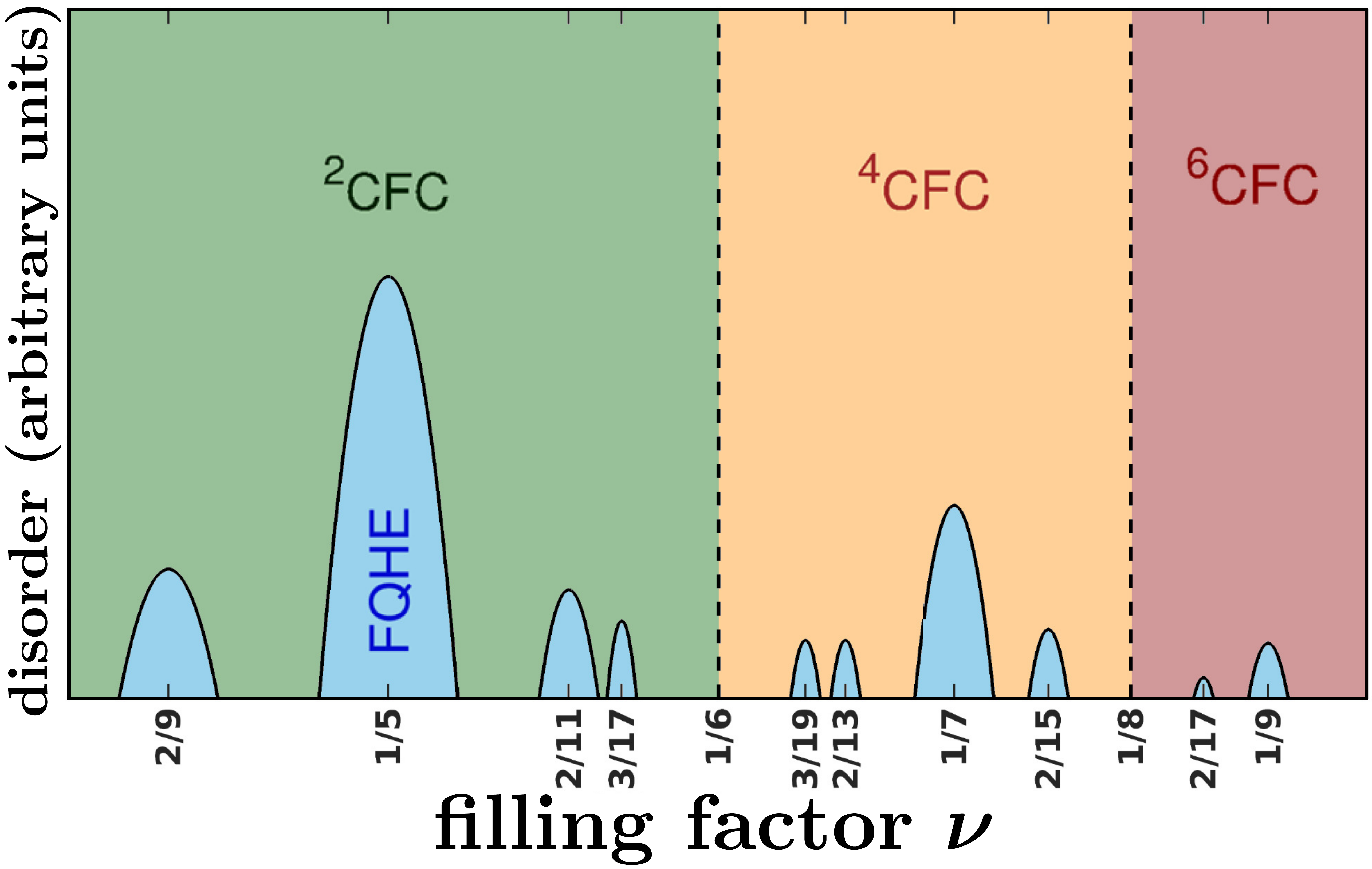} 
\caption{Schematic phase diagram implied by the our conclusions. The acronyms $^{2}$CFC, $^{4}$CFC and $^{6}$CFC, refer respectively to crystals of composite fermions carrying two, four and six vortices.}
\label{phase-diagram}
\end{center}
\end{figure}

One may consider the alternative possibility where the ground state is a crystal for $\nu<1/5$ in the absence of disorder. This crystal is pinned by the disorder to produce insulating behavior. The observation, at somewhat higher temperatures, of resistance minima at the special filling factors would imply that, as the temperature is raised, the crystal melts into a correlated FQH liquid. In other words, electrons must bind additional vortices when the crystal melts. While in principle possible, this appears to us to be counterintuitive, because one expects electrons to shed their vortices as the temperature is raised. 

A definitive experiment to distinguish between the two scenarios would be to determine whether the resistance minima at the special filling factors appear only above certain ``melting" temperature (a pure crystal phase cannot have any FQHE-like signatures) or whether they persist to arbitrarily low temperatures. The difficulty is that at low temperatures the resistance becomes too large to measure. In the published experiments~\cite{Goldman88, Mallett88, Pan02}, however, the minima are seen whenever the resistance is measurable; in fact, the minima do not disappear with lowering the temperature but rather deepen relative to the exponentially rising background. Experiments on better quality graphene or GaAs systems (where substantial progress has recently been made toward achieving higher mobilities~\cite{Chung20,Pfeiffer20}) may shed definitive light on this important question.

%%%%%%%%%%%%%%%%%%%%%%%%%%%%%%%%%%%%%%%%%%%%%%%%%%%%%%%%%%%%%%%%%%%%%%%%%
\begin{acknowledgments} The authors are grateful to M. P. Zaletel, and R. S. K. Mong for their insights on the issue of ground-state energy with DMRG. The work at Penn State (S.P., J.Z., J.K.J.) was supported by the U. S. Department of Energy under Grant No. DE-SC0005042. Some portions of this research were conducted with Advanced CyberInfrastructure computational resources provided by The Institute for CyberScience at The Pennsylvania State University and the Nandadevi supercomputer, which is maintained and supported by the Institute of Mathematical Science’s High-Performance Computing Center. This work was also supported by the Polish NCN Grant No. 2014/14/A/ST3/00654 (A. W.). We thank Wroc\l{}aw Centre for Networking and Supercomputing and Academic Computer Centre CYFRONET, both parts of PL-Grid Infrastructure. Some of the numerical calculations were performed using the DiagHam package, for which we are grateful to its authors. Z.W.Z. was also supported by the National Natural Science Foundation of China under Grants No. 11604081 and No. 11447008. We thank the Tianhe-2 platform at the National Supercomputer Center in Guangzhou for technical support and a generous allocation of CPU time. One of us (Th.J.) acknowledges CEA-DRF for providing CPU time on the supercomputer COBALT at GENCI-CCRT. 
\end{acknowledgments}
$^{\dagger}$Z. W. Z. and A. C. B. contributed equally to this work.

\bibliography{biblio_fqhe} 
\bibliographystyle{apsrev}
\end{document}